# TRIDS: AI-native molecular docking framework for accelerating high-throughput virtual screening with physically valid binding poses

Xuhan Liu[1,8,†], Baohua Zhang[2,6,1,†], Yue Xue[3,7], Yize Hao[3], Tiejun Bing[9], Lili Chai[9], Tanfeng Zhao[9], Lijiang Yang[3], Hong Zhang[4,*], Yi Qin Gao[1,3,4,5,*]

[1]Institute of Systems and Physics Biology, Shenzhen Bay Laboratory, Shenzhen, 518055, China
[2]Shenzhen medical academy of research and translation, Shenzhen, 518055, China
[3]Beijing National Laboratory for Molecular Sciences, New Cornerstone Science Laboratory, College of Chemistry and Molecular Engineering, Peking University, Beijing, 100871, China
[4]Changping Laboratory, Beijing 102200, China
[5]Department of Chemistry, Westlake University, Hangzhou, 310000, China
[6]Westlake University, Hangzhou, 310000, China
[7]Chromatin (Beijing) Technology Co., Ltd., Beijing, China
[8]XtalPi Technologies *Co., Ltd.*, Shenzhen, 518000, China
[9]ICE Bioscience Inc., Bldg 16, Yd 18, Kechuang 13th St, Etown, Tongzhou Dist, Beijing, 100176, China
[†]The first two authors should be regarded as Joint First Authors
[*]To whom correspondence should be addressed.

Email address of authors:
(X.L) xuhan.liu@xtalpi.com
(B.Z) zhangbaohua@smart.org.cn
(H.Z) zhangh@pku.edu.cn
(Y.Q.G) gaoyq@pku.edu.cn

ORCID:
(X.L) 0000-0003-2368-4655
(B.Z) 0009-0007-1474-7841
(H.Z) 0000-0002-3303-5109
(Y.Q.G) 0000-0002-4309-9376

## Abstract

Molecular docking is a cornerstone of drug discovery for unveiling the mechanism of ligand-receptor interactions. With the recent advances of deep learning (DL), AI-powered molecular docking methods have achieved higher accuracy for binding pose prediction and virtual screening compared with classical physics-based methods. However, there is still a scarcity of approaches to strike a balance among accuracy, computational efficiency, and rigorous physical validity of the output conformations. In the previous two versions of DSDP, we demonstrated the effectiveness of guiding conformation sampling with the gradient of an analytic scoring function. As the third version, *TRIDS* was devised as an AI-native docking framework that expand the similar strategy to unify conformation sampling and docking processes with DL-based model for improving accuracy of docking and screening. Furthermore, it is tailored for seamless cooperation of AI and physics to guarantee the physical validity of predicted binding poses. Being user-friendly, *TRIDS* predicts the binding site, parses multiple file formats, and supports Python programming and PyMOL graphical interaction. It improves docking accuracy and passes physical validation with high computational efficiency, *i.e.* a single docking task is done in a fraction of second while maintaining a highly lightweight GPU memory footprint of merely hundreds of megabytes, facilitating high-throughput virtual screening in reality. As a proof of concept, *TRIDS* allowed us to obtain hit compounds with novel scaffolds for tumor necrosis factor-alpha (TNFα) inhibitor through a large-scale virtual screening.

## 1. Introduction

Molecular docking plays a pivotal role in drug discovery by predicting optimal binding conformations and affinities of receptor-ligand interactions[1]. Governed by principles of spatial complementarity and energy minimization, it identifies ligand poses that maximize favorable interactions while minimizing steric clashes and potential energy. It underpins critical drug discovery workflows including virtual screening (VS), polypharmacology, drug repositioning, and off-target prediction[2,3]. In recent years, with the rapid development of machine learning (ML), particularly deep learning, the paradigm of molecular docking has been shifted tremendously in all steps, including binding site identification, conformation sampling, and scoring.

Accurate binding site identification represents the foundational stage of molecular docking, particularly in blind docking scenarios where experimental ligand-bound structures are unavailable. Traditional approaches encompass geometry-based cavity detection, energy-based probe scanning, and template-based methods. For example, COACH[4] integrates binding-specific substructure comparisons, sequence profile alignments, and complementary prediction tools through support vector machine training. The application of these methods, however, is fundamentally constrained by protein conformations and the inherent difficulty in modeling diverse intermolecular interactions including hydrogen bonding, hydrophobic effects, and π-stacking. Deep learning approaches, such as DeepSite[5], discretize the 3D space into grids to predict per-point binding probabilities and generate geometrically adaptive pockets. Their effectiveness remains significantly hampered by insufficient high-quality training data[6]. Hybrid strategies exemplified by P2Rank[7], which synergistically integrates traditional algorithms with deep learning, typically demonstrate superior predictive performance.

Conformation sampling aims at identifying biologically relevant conformations through systematic exploration of ligand poses within predetermined binding sites. It has three notable challenges: inherent system complexity, exponential expansion of sampling space with increasing degrees of freedom, and the competing demands of atomic-level precision versus computational efficiency in high-throughput contexts[8]. Existing sampling methodologies can be broadly classified into traditional and ML-based categories. Traditional approaches are differentiated by their treatment of ligand degrees of freedom[9]. Shape-matching algorithms like Surflex[10] leverage molecular similarity metrics to incrementally construct poses through fragment morphing. Systematic search techniques such as Glide[11] exhaustively enumerate torsional energy minima to generate initial poses, which are subsequently refined through energy minimization and Monte Carlo optimization[12] to ensure robustness. Stochastic methods including AutoDock[13] and MOLDock[14] probabilistically modify poses using predefined acceptance criteria via Monte Carlo or genetic algorithms. Notably, GPU acceleration has dramatically enhanced traditional sampling efficiency. For example, Vina-GPU[15] achieves a more than 60-fold docking acceleration compared to the original CPU version. DSDP[16] implements the gradient calculation of analytic scoring function on GPU to guide conformation sampling. It also combines a ML-based binding site prediction model to achieve high success rates (SR) in blind docking benchmarks and reduces processing time to ~ 1 second per system. ML-based approaches such as EquiBind[17] and DiffDock[18] utilize equivariant neural networks and generative diffusion models to directly generate conformations, respectively. More recently, CarsiDock[19] is trained on the large augmented dataset to predict atomic distance matrix and achieves accurate prediction of ligand conformations. Additionally, SurfDock[20] employs diffusion models based on protein geometric surface features to generate precise and reliable receptor-ligand complex conformations. However, the ML-based methods are prone to generating physically invalid conformations and thus limit their ability to unveil key receptor-ligand interactions, which act as the important hints to reflect the biological relevance for drug discovery[21].

Construction and evaluation of scoring functions (SFs) constitute the third essential task in molecular docking. A qualified SF must meet several critical criteria, including accuracy, computational efficiency, and generalizability, as they directly discriminate how "good" the sampled conformations are and determine the success of VS campaigns. Classical SFs, which typically employ analytic additive formulations, can be categorized into three classes[22]: physics-based methods such as GoldScore[23] often utilize functional forms inspired by classical molecular mechanics force fields; empirical functions like GlideScore[11] employ parameterized energy terms optimized for computational speed yet exhibit pronounced training set dependency; knowledge-based approaches (*e.g.* DrugScore[24]) leverage statistical potentials derived from diverse receptor-ligand complexes to balance accuracy and generalizability.

In contrast, ML-based SFs learn hidden patterns from large datasets of known target-ligand complexes. They often encode these data pair of protein and ligand into physicochemical and geometric descriptors, thereby inferring properties such as binding affinity for unknown complexes. Traditional ML-based methods, such as RF-Score[25], were typically trained with the observed frequencies of receptor-ligand atom pairs as input features for their random forest models. In comparison, deep learning models can process large, multi-modal and heterogeneous data in an end-to-end fashion, *i.e.*, by directly mapping raw input data to final output predictions. It can be further categorized into supervised and unsupervised learning approaches. GNINA[26], for example, is a supervised SF based on 3D CNNs, whereas RTMScore[27] is an unsupervised GNN-based method. In spite of their flexibility, ML-based SFs are also subject to several limitations, including poor interpretability[28], insufficient accounting for solvent and entropic effects[29], and data scarcity[30], particularly for supervised methods requiring extensive affinity data augmentation. Consequently, while classical SFs dominate current production pipelines, ML-based approaches demonstrate growing promise for specialized applications despite their inherent constraints.

Despite aforementioned representative works, molecular docking methods are still faced with major challenges in accuracy, efficiency and generalizability. Although the advancement of deep learning has introduced novel paradigms for molecular docking, few existing methodologies successfully reconcile the competing demands of high accuracy, computational efficiency, and strict physical plausibility in the resulting conformations.[31] To overcome some of the difficulties and inherited from our previous two versions of DSDP[16,32], we expanded the idea of conformation sampling guided by the gradient of differentiable DL-based SFs to devise the third version. As an AI-native docking framework, it was implemented with *PyTorch C++* (*LibTorch*) to unify of Deep learning models for binding Site prediction, Scoring and Sampling (named *TRIDS*). Moreover, it is tailored for seamless cooperation of AI and physics to guarantee the physical validity of predicted binding poses. Comprehensive tests and experimental validation were performed to evaluate the docking accuracy and computational efficiency of *TRIDS*.

## 2. Methods and Materials

**Dataset preparation**

During the process of model training, BioLip[33] and PDBbind (v2020)[34] were used for pretraining and fine-tuning, respectively. In addition, a variety of benchmark sets were employed in this study to comprehensively evaluate our docking framework, including CASF-2016[35] for the benchmark of the performance of different SFs, PDBbind time-split set (Time-split)[17] and PoseBusters[36] for the estimation of the docking power, and DEKOIS 2.0[37] for the test of the screening power.

In this work, a two-stage training strategy was adopted: 1), large-scale pretraining on a diverse set of receptor-ligand complexes to learn general interaction patterns, and 2) task-specific fine-tuning on high-quality datasets. For the pretraining process, non-covalent small molecules from the BioLiP were utilized; after data cleaning, approximately 170,000 molecules remained. For the fine-tuning, the PDBbind was utilized here. The receptor-ligand complexes in CASF-2016[35], Time-split and PoseBusters, as well as those not processed by OpenBabel[38], are eliminated in both pretraining and fine-tuning processes. 1,500 entries were randomly selected for validation and the rest were used for training. The validation set here was employed for the judgment of early stopping in model training to avoid over-fitting as well as for the selection of the model that exhibited optimal performance. To balance between accuracy and computational cost, residues with atomic distances less than 8Å[39] to the reference ligand were extracted as binding pocket of receptors. In the following process of scoring and sampling, only residues in the binding pocket were referred to in the computation. In the absence of reference ligand, the binding site prediction model, the architecture and parameters of which were the same as DSDP, was invoked automatically during the inference process. Two models were trained in the present work, namely, a redocking model pretrained on BioLip and fine-tuned on PDBbind, and a screening model pretrained on BioLip and fine-tuned on the pocket augmentation set of PDBbind. For the screening task, we adopted the pocket augmentation strategy similar to CarsiDock[19]. Specifically, for each protein, two initial pockets were generated using cutoffs of 7.0 Å and 9.0 Å around the co-crystallized ligand. The residues that differ between these two pockets were then randomly added to the 7.0 Å pocket, resulting in 10 augmented pockets with diverse shapes. Ultimately, we expanded the final pocket set tenfold based on the PDBbind database.

### Scoring functions

The main SF in the present work is the mixture density network (MDN) derived from RTMScore[27], the complete framework is shown in Fig. 1. Firstly, features of input protein and ligand are computed separately using OpenBabel[38], and then graph structures at the residue and atom levels are constructed using the PyTorch Geometric (PyG) package[40] (Fig. 1A). Tables S1 and S2 summarize the input node and edge features for the ligand and protein graphs, respectively.

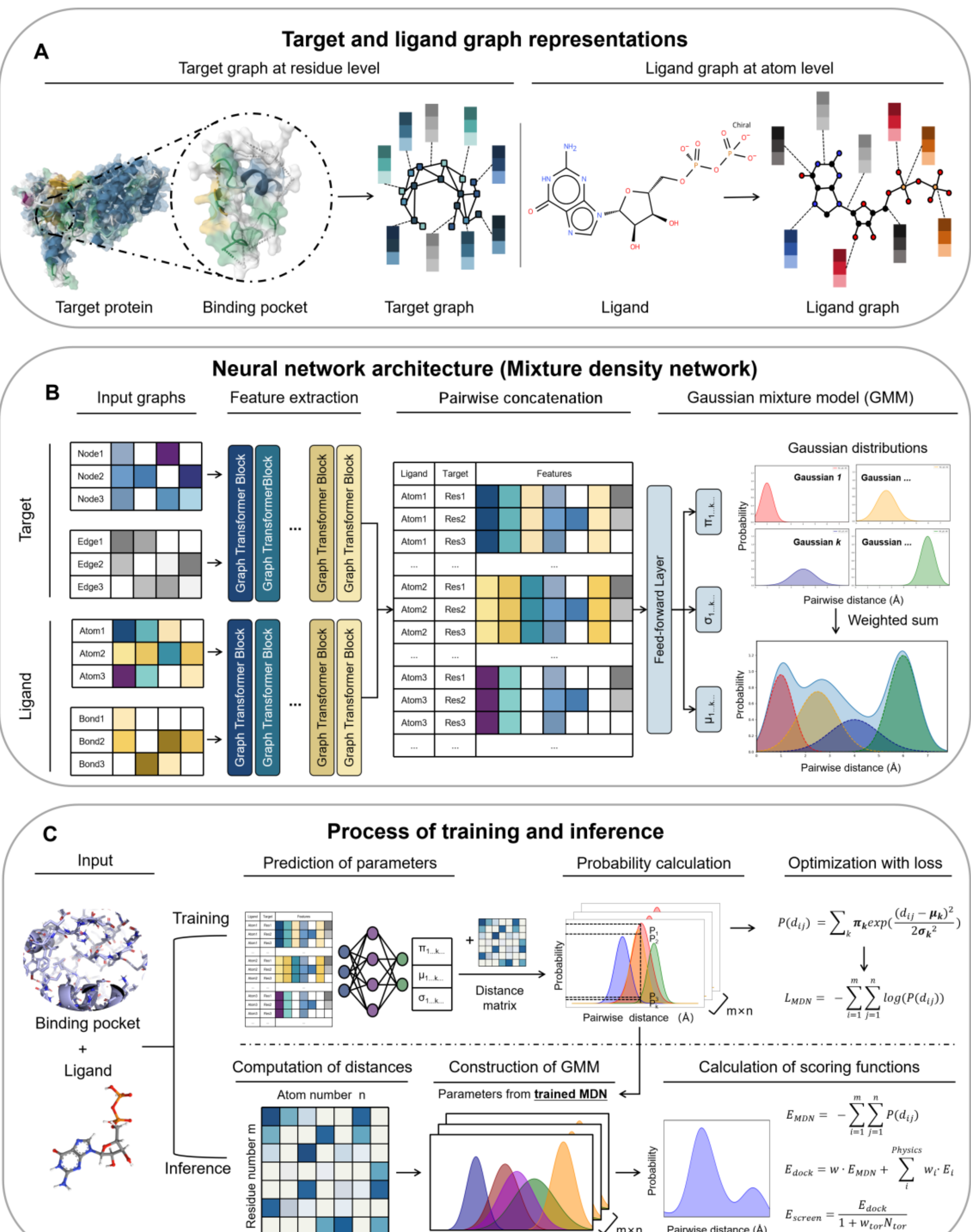

**Fig. 1:** The framework of the *MDN-based SF* used in *TRIDS*. (A) Graph representations of protein and ligand in *MDN*. Each receptor and ligand is represented as a residue-level graph with spatial proximity and atom-level graph connected by covalent bonds, respectively. (B) Architecture of MDN-based neural networks. The graph representations are processed by Graph Transformer blocks to extract node features. They are then pairwise-concatenated and fed into *MDN* that parameterize GMM for probability distributions with respect to pairwise distances. (C) Training and inference processes of *MDN*. The model is trained through minimization of the negative log-likelihood function. The well-trained model takes each receptor-ligand complex as input to predict the profile of GMM, with which the score for docking or screening could be inferred based on the distance matrix.

The architecture of MDN consists of three key components: a feature extraction module, a feature concatenation module, and a feed-forward output layer (Fig. 1B). The graphs obtained from the previous step are processed by multiple Graph Transformer, so that the processed node features contain not only information about an individual atom or residue in the molecule, but also information about the surrounding nodes. The resulting graphs are then pairwise-concatenated to model the interaction and fed into a feed-forward layer to parameterize a GMM, which contained multiple weighted Gaussian kernels to measure the probability with refer to distance of atom-residue pair within the pre-defined binding pocket:

$$P(d_{ij}) = \sum_{k} \pi_k \cdot e^{\frac{(d_{ij}-\mu_k)^2}{2\sigma_k^2}}$$

where $d_{ij}$ denotes the minimum distance between the atom in protein residue $i$ and ligand atom $j$; $\sigma_k$, $\mu_k$ and $\pi_k$ are the mean, standard deviation and weight parameters in the $k^{th}$ Gaussian kernel predicted by MDN.

During the training phase, a pairwise distance matrix between the $m$ atoms of the ligand and the $n$ residues of the protein is first computed. Subsequently, the probability of each node pair is calculated based on the GMM, and the sum of the negative log-likelihoods is used as the MDN loss function for back-propagation optimization:

$$L_{MDN} = -\sum_{i=1}^{m}\sum_{j=1}^{n} (d_{ij} < d_{cutoff})\ log(P(d_{ij}))$$

It should be noticed that only node pairs with distances less than $d_{cutoff}$= 5.0 Å are used in training process. The models were optimized using the Adam optimizer, with a batch size of 32, a learning rate of $10^{-3}$ and $10^{-5}$ for pretraining and fine-tuning, respectively. With a weight decay of $10^{-5}$, training process was halted early if the validation loss did not improve over 100 consecutive epochs.

During the inference phase (Fig. 1C), for each receptor-ligand pair, the well-trained MDN firstly predicted the parameters in GMM, and then the binding score ($E_{MDN}$) was computed with the distance matrix of atoms between the given receptor and ligand. The formula of $E_{MDN}$ is given below:

$$E_{MDN} = -\sum_{i}^{m}\sum_{j}^{n} (d_{ij} < d_{cutoff})\, P\left(d_{ij} \mid h_i^{rec}, h_j^{lig}\right)$$

where $h_i^{prot}$ and $h_j^{lig}$ are extracted graph features for receptors and ligands, respectively. Similar to training process, only node pairs with their distances smaller than $d_{cutoff}$= 5.0 Å are involved in the inference process.

In order to improve the docking accuracy, especially the physics-based validity, three physics-based terms were integrated into our SF, including:

1) long range interaction term, which is defined as Gaussian formula originated from the second Gaussian term in VINA:

$$E_{long} = -\sum_{i}\sum_{j}\left(d_{ij} < d_{max}\right) e^{\frac{(d_{ij}-d_{vdw}-\mu)^2}{2\sigma^2}}$$

where $\sigma$ and $\mu$ are Gaussian coefficients as in VINA, $d_{vdw}$ is sum of Van der Waals radius of the given atom pair, and $d_{max}$ is set as 10 Å;

2) clash penalty term, which is calculated as follows:

$$E_{clash} = -\sum_{i}\sum_{j}\left(d_{ij} < d_{vdw}\right) \log\left(\frac{d_{ij}}{d_{clash}}\right)$$

where $d_{clash}$ is 2.6 if they are the acceptor/donor pair of hydrogen bond, otherwise, equals to $d_{vdw}$;

3) hydrogen bond revision term, which is also given in Gaussian formulae only for those atom pairs of hydrogen bond acceptor/donor:

$$E_{Hbond} = -\sum_{i}\sum_{j}\left(d_{ij} < d_{Hmax}\right) e^{\frac{(d_{ij}-\mu)^2}{2\sigma^2}}$$

where $\mu$, $\sigma$ and $d_{Hmax}$ equals 2.88, 0.3 and 3.4, respectively.

In combination with these three terms linearly, the total score ($E_{total}$) for ranking the docking poses was defined as their summation:

$$E_{dock} = w_{MDN} \cdot E_{MDN} + w_{long} \cdot E_{long} + w_{clash} \cdot E_{clash} + w_{Hbond} \cdot E_{Hbond}$$

with the four weight coefficients ($w_{MDN}$, $w_{long}$, $w_{clash}$, $w_{Hbond}$) set to be 1.0, 0.2, 10.0, and 0.6, respectively. Because all of these terms are calculated based on pairwise distances, the gradient of this SF can be passed to the coordinates of ligand atoms for conformation sampling.

For VS, a revised torsion bond term was also added to filter out irregular ligands with long flexible chains. It was also defined as in VINA:

$$E_{screen} = \frac{E_{dock}}{1 + w_{tor} \cdot N_{tor}}$$

where $N_{tor}$ was number of rotatable bonds involved in the sampling process, and the weight coefficient $w_{tor}$ equals to 0.01.

**Conformation sampling**

The sampling space is defined on the degrees of freedom for each molecule, including translation, rotation and torsion angles of rotatable bonds, the ranges of which are [$box^{min}$, $box^{max}$], [$-\pi$, $\pi$), and [$-\pi$, $\pi$), respectively. Here, $box^{min}$, $box^{max}$ are minimum and maximum of box lengths determined by the reference ligand, which is conformed to interact with binding pocket of the given receptor. The objective of conformation sampling is to optimize these variables to achieve the best score judged by the SF. Here, the combination of Metropolis Monte Carlo and gradient descent algorithms are adopted as the global minimization strategy for conformation sampling (Fig. 2). The number of CUDA streams and depths were set to 1,024 and 8, respectively, *i.e.*, a batch of conformation with 1,024 copies were run simultaneously, each of which was individually tackled with single CUDA stream for eight computational loops in total, including perturbation, refinement and acceptance validation. In the beginning, each copy of conformation was initiated with randomization of the degrees of freedom involved. During the computational loop, one variable was randomly perturbed to update the conformation at first. Subsequently, the updated conformation was optimized through the gradient descent method. Adam with learning rate 0.1 was used to update the variables, in which the above-mentioned ML-based scoring model was differentiated to provide gradients. At the end of each loop, the perturbation was accepted according to the following criterion:

$$\text{rand}\ (0,\ 1) < \exp(\ -\beta\ \Delta E)$$

In the above equation, $\beta$ is a hyper-parameter used in the Monte Carlo sampling process, and the value of $\beta$ is set to 0.069 for the sake of making the probability of acceptance equals 0.5 for every difference of 10 points calculated by SF ($\Delta E$). Subsequently, these samples were ranked and clustered to filter out duplicates. Finally, top $N$ conformations were selected as the predicted binding poses.

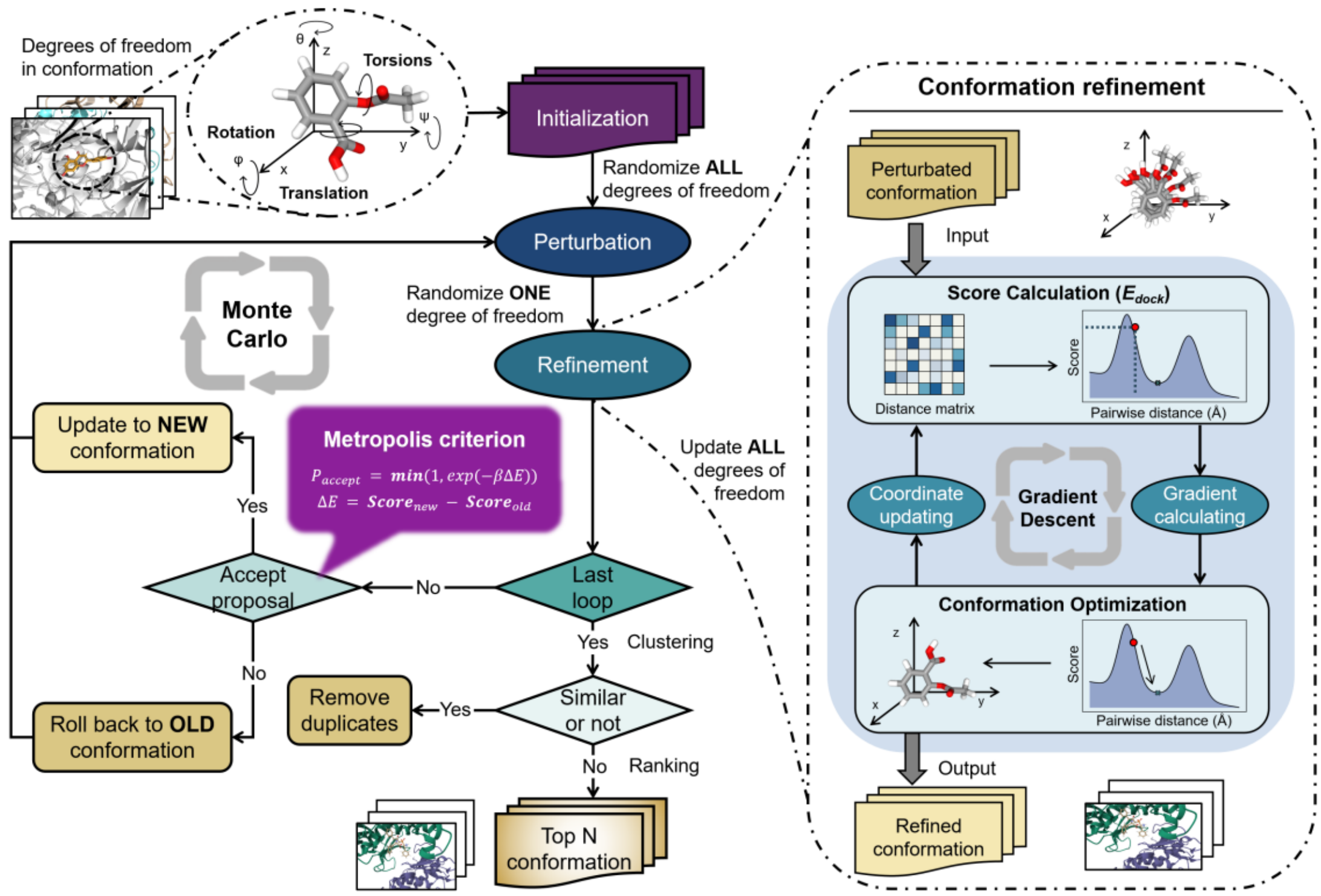

**Fig. 2:** The flowchart of conformation sampling in *TRIDS* based on Metropolis Monte Carlo algorithm, including conformation initialization, perturbation, refinement and proposal acceptance. Here, the refinement of conformation was implemented by gradient descent method in which the gradient of both MDN-based SF and additive physics-informed terms was calculated for guiding the conformation optimization.

**Model evaluation**

Firstly, CASF-2016[41] was used in the initial validation of our SF in the docking and screening tasks. Docking performance was evaluated based on the SR, where a prediction is considered successful if the root-mean-square-deviation (RMSD) between the best-scored pose and the native structure is below a predefined threshold (typically 2.0 Å). Screening power was assessed using the SR for identifying the highest-affinity binder among the top 1% ranked ligands in "forward screening", as well as the enrichment factor (EF), defined as the ratio of true binders found within the top 1% of ranked compounds.

Secondly, Time-split was used to evaluate the generalizability of different methods, in which complexes released in 2019 or later were used as the test set, and earlier structures for training and validation. Performance was measured based on SR among the top 1 predicted conformation and the median of their RMSD.

Subsequently, the PoseBusters[36] developed recently was also employed here to further test the docking accuracy of our approach. This dataset consists of 428 receptor-ligand crystal structures published from 2021 onward, ensuring no overlap with the PDBbind training set, and was used by many DL-guided docking approaches. Besides the conventional SR under the 2.0 Å RMSD threshold, PoseBusters introduces the “PB-valid” criteria (molecule poses which pass all tests in PoseBusters), which consist of chemical validity, intramolecular properties, and intermolecular interactions.

Finally, DEKOIS 2.0[37] was used for evaluation of VS, which includes 81 targets each with 40 active ligands and 1200 decoys. Metrics used in these comparisons include the area under the receiver operating characteristic curve (AUROC), and EF values at different percentiles (0.5% and 1%).

**Virtual screening and experimental assay**

A proprietary database comprising approximately 20,000 virtual small molecules was subjected to molecular docking screening using *TRIDS*. Based on the docking scores, the top 2,000 compounds were initially selected for further analysis. These compounds were then clustered to maintain structural diversity. Following clustering, a property-based filtering and analysis were performed to evaluate drug-likeness and physicochemical characteristics. Based on the combined docking scores and filtering results, a subset of 50 compounds was purchased and procured for subsequent experimental validation. TNFα binding affinity assay was conducted using Dianthus high-throughput affinity screening system, which is based on Spectral Shift (SpS) technology. The assay was performed in a buffer consisting of 20 mM HEPES, 150 mM NaCl, and 0.02% Tween-20. The final reaction concentrations were 25 nM for TNFα protein and 10 nM for the His-Dye. The protein and dye were thoroughly mixed and incubated for 10 minutes to allow complex formation. For the measurement, 19.8 μL of the pre-incubated protein-dye complex was mixed with 0.2 μL of the compound solution. The plates were sealed and incubated at 3000 rpm for 2 hours at room temperature. Fluorescence data were collected using Dianthus instrument. The positive control, TNF-α-IN-2 and Balinatunfib, were tested at a concentration of 10 μM with a 3-fold serial dilution across 12 doses. The estimated compounds identified from the VS were tested at a concentration of 100 μM with a 3-fold serial dilution across 12 doses.

## 3. Results

First of all, the performance of MDN in this work has been tested on CASF-2016. To make a fair evaluation of the performance, more than 40 SFs referred to previous work[42,43] were used for comparison, including representative classical and ML-based SFs. These results showed that our constructed MDN was equivalent to that of RTMScore, which indicates that it is a competent SF for the docking and screening tasks (Fig. S1). Subsequently, the well-trained MDN was integrated into *TRIDS* for the evaluation of docking and screening jobs, to compare with several state-of-art molecular docking methods. For PDBbind times-split (Time-split) and PoseBusters, docking results were directly retrieved from several molecular docking methods for comparison, including AutoDock Vina[13], TANKBind[44], Equibind[17], DiffDock[18], GNINA[26], CarsiDock[19], and SurfDock[20]. Among these methods, Equibind, TANKBind, and DiffDock are typical ML-based docking methods. AutoDock Vina and its derivate GNINA are commonly used traditional programs. CarsiDock and SurfDock, both of which apply ML-based sampling methods and exhibited high accuracy in docking tasks, are utilized as the main baselines to be compared with *TRIDS*. For DEKOIS 2.0, our method was compared with representative methods of VS, which included TANKBind, KarmaDock, Vina, GNINA, Surflex-Dock[10], Glide SP[11], CarsiDock[19] and SurfDock[20].

### 3.1 Docking power of *TRIDS*

In docking accuracy evaluation, we placed emphasis on the overall performance, *i.e.* achieving both high docking accuracy and maintaining physical validity. To validate whether the conformations obtained by various docking methods meet physical validation criteria, all docking results were evaluated against the physical criteria defined by PoseBusters. The docking on Time-split (Table 1) shows that the vast majority of successfully docked poses meet the physical validation standards for both GNINA (43.62% SR/41.41% PB Valid) and Vina (36.64% SR/32.87% PB Valid), as both utilize traditional SFs to guide sampling. Although SurfDock achieved high docking accuracy (68.41%), also outperforming other ML-based docking programs such as EquiBind, TANKBind, DiffDock, but a smaller portion of their generated conformations met physical validation benchmarks (36.46%) compared to traditional methods. This result reveals the challenge for ML-based methods to take the strict physical fitness into consideration alongside docking accuracy assessment. The docking power of *TRIDS* (66.94%) was much higher than traditional physics-based approaches. Although this SR is lower than that of SurfDock, *TRIDS* achieved the highest proportion (57.83%) of physically valid poses, which is even more effective than SurfDock followed by energy minimization using molecular dynamics simulations.

**Table 1:** Docking performance of *TRIDS* compared with other docking methods evaluated on Time-split

| Method | Description of the method | Top-1 RMSD | | |
|---|---|---|---|---|
| | | <2 Å (%) ↑ | Med. (Å) ↓ | <2 Å & PB Valid[b] ↑ |
| EquiBind[a] | DL sampling | 5.5 | 6.2 | / |
| TANKBind[a] | DL sampling | 18.18 | 4.2 | / |
| DiffDock[a] | DL sampling | 36.09 | 3.35 | 15.43 |
| AutoDock | Classical sampling | 36.64 | 3.42 | 32.87 |
| GNINA[a] | Classical sampling (rescoring) | 43.62 | 2.45 | 41.41 |
| DSDP | Classical sampling | 49.73 | 2.04 | / |
| CarsiDock[c] | DL sampling | 66.30 | / | / |
| SurfDock[a] | DL sampling (rescoring) | **68.41** | 1.18 | 36.46 |
| SurfDock[a] (minimized) | DL sampling (rescoring + energy minimization) | 68.04 | **1.1** | 55.0 |
| TRIDS | DL sampling | 66.94 | 1.23 | **57.83** |

[a]: these results are from SurfDock work by Cao et al[20].

[b]: the systems failed recognized by bust were removed to estimate the SR.

[c]: Results from Ref.[19]

Next, PoseBusters was used as an unbiased dataset to further estimate the performance of *TRIDS*. It comprises of 25 distinct physical metrics. We observed that *TRIDS* passed criteria for 23 of these metrics. These successfully validated metrics were therefore excluded from visualization in Fig. S2, which showed that the minimum distance between atoms of ligand and protein constituted the dominant source of failures in physical validation. In addition, the SR of *TRIDS* was 77.3% in the redocking task, lower than SurfDock and Carsidock, but the PB_Valid proportion (75.9%) of *TRIDS* surpassed all other methods (Fig. 3A), demonstrating its effectiveness in prediction of physics-consistent binding poses. To evaluate the generalization capability of docking methods across diverse protein systems, PoseBusters categorized targets into three ranges: 0-30%, 30-95%, and 95-100% according to their sequence similarity with PDBbind. Again, *TRIDS* performs similarly to CarsiDock and SurfDock, exhibiting no strong correlation between performance and sequence similarity (Fig. 3B). No significant loss in docking accuracy was observed as sequence similarity decreased.

Furthermore, we examined how the docking accuracy is affected by the molecular weight on PoseBusters. The SR of SurfDock significantly decreased along with the increase of molecular weight (Fig. 3C), and the median of RMSD (Fig. 3D) also increased sharply when the molecular weight is higher than 500 Da. These results show that the docking accuracy of SurfDock is limited for large molecules. A similar but weaker trend is also found for CarsiDock. In contrast, the accuracy and the distribution of RMSD of *TRIDS* decay slower with molecular weight than the other two methods. Since the average molecular weight of the approved drugs in the past five years has exceeded 500 Da, the capability of *TRIDS* in handling large molecules makes it particularly valuable for drug screening..

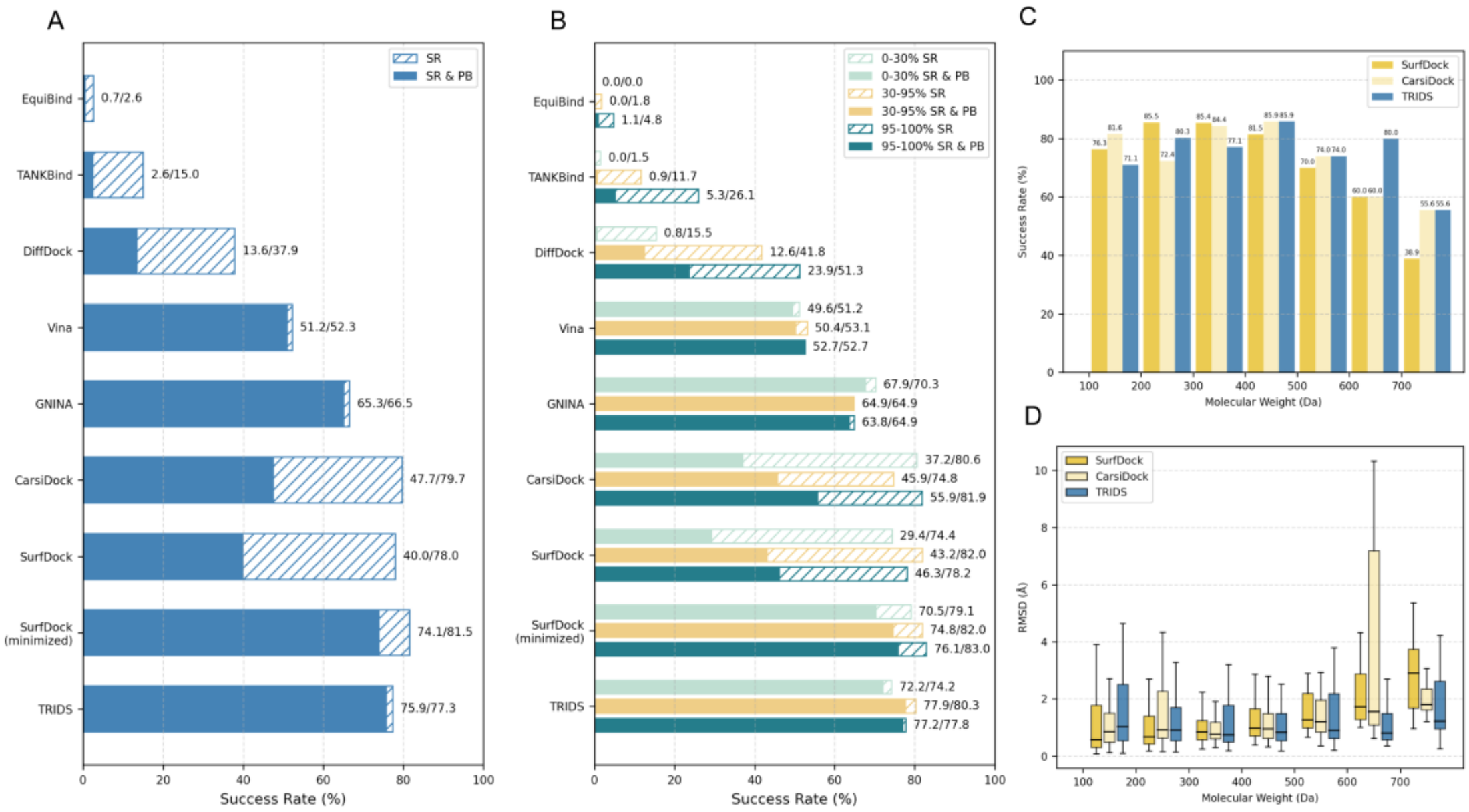

**Fig. 3:** Performance benchmark of *TRIDS* and other methods on the PoseBusters, including (A) the SR of docking powers, (B) the SR of docking powers categorized by sequence similarity to PDBbind, (C) SR and (D) RMSD Median of SurfDock, CarsiDock, and *TRIDS* along with the molecular weight in PoseBusters. The results of other methods are from work SurfDock[20] and CarsiDock[19].

### 3.2 Screening power of *TRIDS*

To rigorously evaluate the efficacy of *TRIDS*, we conducted a comprehensive benchmarking analysis on the DEKOIS 2.0, a standard benchmark for assessing virtual screening (VS) performance. As summarized in Table 2, *TRIDS* demonstrates strong predictive capability, achieving the highest ROC-AUC (0.815) among all tested protocols. Furthermore, it shows excellent early screening performance, achieving the highest EF0.5% (21.30) and EF1% (19.43). These results provide compelling evidence that the unification of sampling and scoring processes within our framework significantly enhances predictive accuracy. Consequently, this integrated strategy exhibits robust performance and high reliability, making it particularly suitable for high-throughput VS tasks where the efficient identification of active compounds from large libraries is critical.

**Table 2:** Performance of *TRIDS* compared with other docking methods for VS tasks on DEKOIS 2.0

| **Method**[a] | **ROC-AUC** | **EF0.5%** | **EF1%** |
| --- | --- | --- | --- |
| TANKBind | 0.60 | 2.83 | 2.90 |
| AutoDock Vina | 0.63 | 5.46 | 4.51 |
| Surflex-Dock | 0.673 | 8.36 | 7.30 |
| GNINA | 0.686 | 11.67 | 9.81 |
| Glide SP | 0.747 | 14.61 | 12.47 |
| KarmaDock (align) | 0.744 | 16.78 | 15.2 |
| CarsiDock | 0.793 | 20.46 | 18.91 |
| SurfDock | 0.758 | 21.00 | 18.17 |
| TRIDS | **0.815** | **21.30** | **19.43** |

[a]The results except GNINA and *TRIDS* are directly retrieved from previous study.[19,20]

### 3.3 Ablation study

The improved performance of *TRIDS* is largely attributed to the pretraining with data augmentation and incorporation of physical energy terms in the sampling. To verify the effectiveness of these strategies, we conducted a few ablation studies. Specifically, we first evaluated the contribution of each energy term (clash, long-range interactions, and hydrogen bonds) to docking accuracy on Time-split (Table 3). Next, we assessed the impacts of the two data augmentation strategies (pre-training on BioLip and protein pocket conformation augmentation) on screening performance using the DEKOIS 2.0 (Table 4).

As shown in Table 3, for the model trained only on PDBbind (Model 1), the $E_{MDN}$ baseline achieved an SR of 39.50%. Adding the clash term, long-range interactions, and hydrogen bonds sequentially increased the SR to 58.93%, 64.00%, and 64.92%, respectively, while the PB_valid proportion improved from 19.33% to 56.63%. The pre-trained model (Model 2) exhibited similar progressive gains (from 39.69% to 66.94%; PB_valid proportion 57.83%). Notably, under the full combination of energy terms, Model 2 outperformed Model 1 in SR (66.94% vs. 64.92%), median RMSD (1.23 Å vs. 1.29 Å), and PB_valid proportion (57.83% vs. 56.63%). It demonstrates that pre-training on BioLip followed by fine-tuning further improves docking accuracy, making Model 2 more suitable for docking tasks. In addition, we also compared the performance of different optimizers for gradient descent such as SGD, AdamW, AdaGrad, and RMSprop. In the end, Adam was proved to perform better than others (Table S5).

**Table 3:** Impacts of several strategies on the docking accuracy on Time-split

| | | | Top-1 RMSD | | |
|---|---|---|---|---|---|
| Model | Training strategy[a] | Terms of SFs[a] | <2 Å (%) ↑ | Med. (Å) ↓ | <2 Å &PB Valid[b] |
| Model 1 | Training on PDBbind only | $E_{MDN}$ | 39.50 | 3.20 | 19.33 |
| | | $E_{MDN}+E_{clash}$ | 58.93 | 1.54 | 51.75 |
| | | $E_{MDN}+E_{clash}+E_{long}$ | 64.00 | 1.31 | 55.16 |
| | | $E_{MDN}+E_{clash}+E_{long}+E_{Hbond}$ | 64.92 | 1.29 | 56.63 |
| Model 2 | Pretraining on BioLip and fine-tuning on PDBbind | $E_{MDN}$ | 39.69 | 2.95 | 20.54 |
| | | $E_{MDN}+E_{clash}$ | 59.48 | 1.43 | 51.75 |
| | | $E_{MDN}+E_{clash}+E_{long}$ | 65.38 | 1.29 | 56.17 |
| | | $E_{MDN}+E_{clash}+E_{long}+E_{Hbond}$ | 66.94 | 1.23 | 57.83 |

[a]All strategies are tested on protein without water.

For the VS task (Table 4), the model trained only on PDBbind (Model 1) achieved EF0.5% of 21.02 and EF1% of 19.05 on DEKOIS 2.0. With BioLip pre-training (Model 2), the EF values were 20.30 and 18.39, respectively. Adding pocket conformation augmentation (Model 3) yielded values of 21.30 (EF0.5%) and 19.43 (EF1%). Model 3 yielded the highest EF0.5% among the three, indicating that pocket conformation augmentation helps improve early enrichment in VS scenarios. In the screening task, the number of rotatable bonds was also involved in the revision of screening score. The comparison of different weights was shown in Table S6, and the weight value was set as 0.01 to balance the performance of ROC-AUC, EF and quantitative estimate of drug-likeness (QED) of selected top 1% ligands.

**Table 4:** Impacts of several strategies on the screening performance on DEKOIS 2.0

| Model | Training strategy[a] | **ROC-AUC** | **EF0.5%** | **EF1%** |
|---|---|---|---|---|
| Model 1 | Training on PDBbind only | 0.795 | 21.02 | 19.05 |
| Model 2 | Pretraining on BioLip and fine-tuning on PDBbind | 0.816 | 20.30 | 18.39 |
| Model 3 | Pretraining with pocket augmentation and fine-tuning on PDBbind | 0.815 | 21.30 | 19.43 |

[a]All strategies and energy terms are tested on protein without water.

[b]the systems failed recognized by bust were removed to estimate the SR.

### 3.4 User-friendliness of *TRIDS*

For the sake of evaluating user-friendliness, we assessed the file parsing SR, GPU memory consumption and docking speed of the different GPU-supported docking methods on Time-split (Table 5). The SR of file parsing was 100% for *TRIDS*, thanks to a full integration with OpenBabel. In contrast, a considerable fraction of files (16.25%) failed to be parsed in CarsiDock and SurfDock, because RDkit is invoked for parsing files and carrying out featurization of molecules. The GPU memory consumption of *TRIDS* was ~380 M, which is 10 times lower than CarsiDock and GNINA, 30 times lower than Vina-GPU, and 100 times lower than SurfDock. The docking speed of *TRIDS* has also advantage over other docking programs using GPU. In the batch docking task, *TRIDS* achieves a molecular docking throughput of 0.33 seconds per receptor-ligand pair on RTX A6000.

Judging by these metrics, *TRIDS* significantly increases the GPU utilization and reduced memory usage, providing a portable platform.

**Table 5:** Software performance of *TRIDS* compared with different types of GPU-supported docking software evaluated on Time-split

| Method | SR of file parsing | Supported file formats | GPU memory | Running Time[b] | Testing Device |
|---|---|---|---|---|---|
| DSDP | 100% | pdbqt | ~800M | 1.2s (0.5s) | RTX A6000 |
| Vina-GPU | 100% | pdbqt | ~10000M | 6s | RTX A6000 |
| GNINA | 100% | sdf, mol2, SMILES, pdb, pdbqt… (all formats supported in OpenBabel) | ~2300M | 62s | RTX A6000 |
| CarsiDock | 83.75% | sdf, mol2, SMILES, pdb (all formats supported in RDkit) | ~3000M | 14s (1.2s) | RTX A6000 |
| SurfDock | 83.75% | sdf, mol2, SMILES, pdb (all formats supported in RDkit) | ~37000M | 115s (3.3s[a]) | A100 |
| TRIDS | 100% | sdf, mol2, SMILES, pdb, pdbqt… (all formats supported in OpenBabel) | **~380M** | 2.3s (**0.33s**) | RTX A6000 |

[a]The docking speed was obtained from the Ref[20],tested on a single H800 GPU (80GB).

[b]The value in parentheses was tested in the batch docking mode on DEKOIS 2.0.

Because of the AI-native framework of *TRIDS*, it could integrate readily with other DL-based models to further enhance functionality. Currently, we have integrated binding site prediction model into *TRIDS*, which has the same architecture and parameters with DSDP. In another word, *TRIDS* also has the capability for blind docking, whose performance on PoseBusters is provided in Table S7. We have engineered a dedicated PyMOL[45] plugin, which allows researchers to conduct binding site prediction (Fig. 4A), molecular docking and scoring (Fig. 4B) with both command line interface (CLI) and graphical user interface (GUI) inside the main window of PyMOL, eliminating complex external software switching. By streamlining the workflow, this plugin significantly improves interactive effectiveness, making advanced docking capabilities readily available for structural analysis in drug discovery applications.

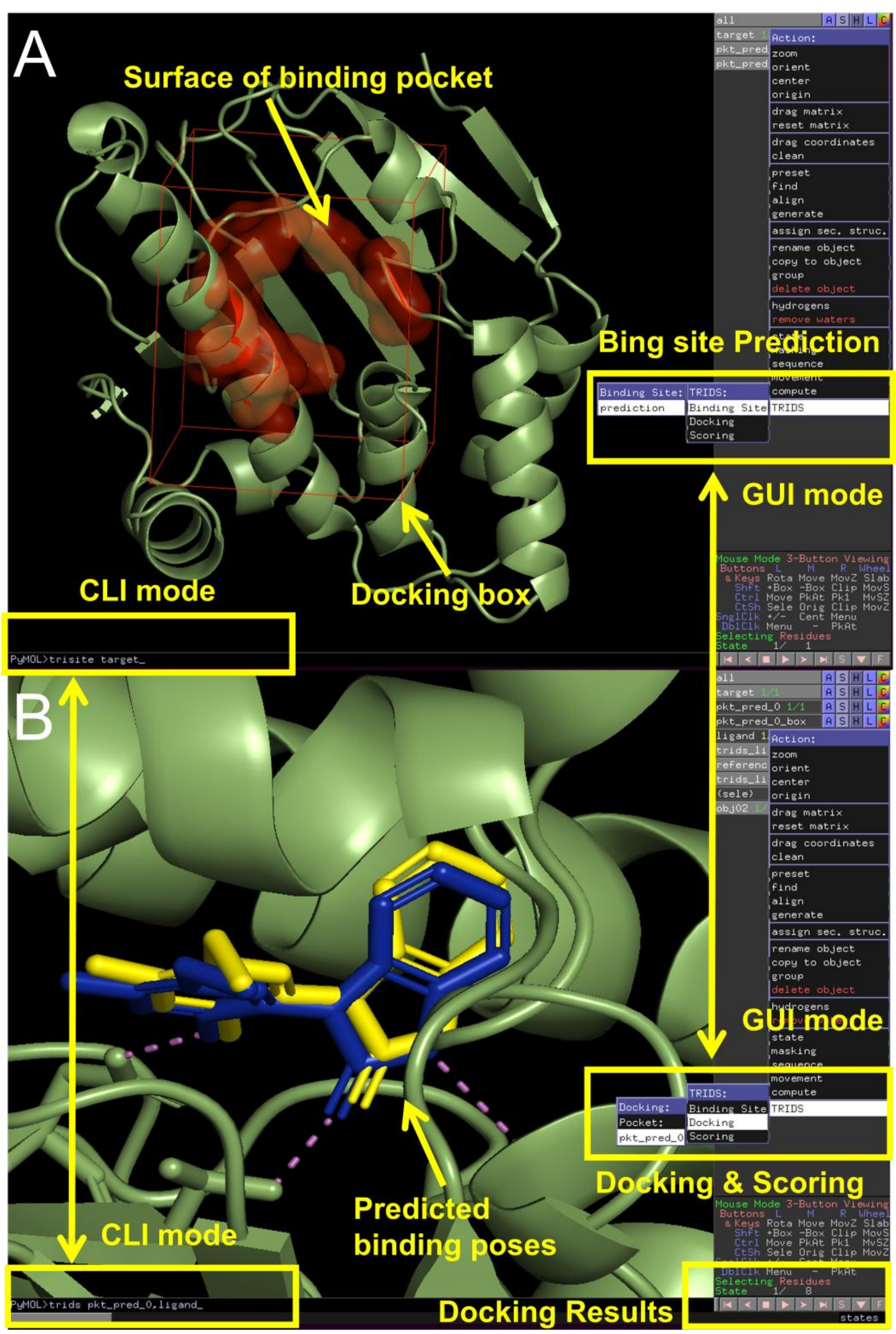

**Fig. 4:** Usage of *TRIDS* plugin inside the main window of PyMOL, including CLI mode and GUI mode. (A) Prediction of the binding site based on the target protein. All of residues closed to the binding pocket are extracted to be a new object surrounded by a cubic box (red). (B) Docking mode of *TRIDS* in PyMOL. There are eight output binding poses (blue) by default compared with the reference ligand (yellow), and these conformations could be found in the state plate.

### 3.5 Experimental validation on TNFα

To further evaluate the practical utility of *TRIDS* in real-world VS tasks, Tumor Necrosis Factor-alpha (TNFα) protein was selected as a target to screening its inhibitors. TNFα is a potent pro-inflammatory cytokine that plays a pivotal role in the pathogenesis of various autoimmune and inflammatory diseases. Biologically active TNFα exists primarily as a homotrimer that exerts its effects by binding to two distinct cell surface receptors, TNFR1 and TNFR2. Inhibiting the interaction between TNFα and its receptor proteins can effectively block the transmission of disease-related signals. There have been reports indicating that small molecules targeting the central site of the trimeric protein can disrupt the active structure of the trimer, effectively inhibiting its correct binding with the receptor protein and thereby blocking the transmission of downstream signals. For this system, we conducted a VS and carried out experimental verification.

A virtual screening task was carried out using approximately 20,000 molecules on TNFα target, and 50 molecules were selected to estimate the binding affinity to the target. The wet experimental screening successfully identified three active molecules out of the 50 molecules (Fig. 5). The dissociation constants (Kd) of these hit compounds were determined to be 40.5, 60.6, and 90.7 μM, respectively. For context, the positive control compounds exhibited activities in the nanomolar range, namely, 374 nM for TNF-α-IN-2 and 905 nM for Balinatunfib. The screened molecules were tens to hundreds of times less potent than the positive controls. However, structural analysis revealed that two (mol_6 and mol_11) of the three identified hits possess novel molecular scaffolds, which are distinct from previously reported TNFα inhibitors. (The molecule mol_39 obtained from VS presents a similar shape as TNF-α-IN-2.) These novel chemotypes thus provide valuable starting points for structural optimization and can contribute to the diversification of lead candidates in TNFα targeted drug discovery. This result indicates that *TRIDS* can reproduce the pivotal binding mode between TNFα and positive control ligand.

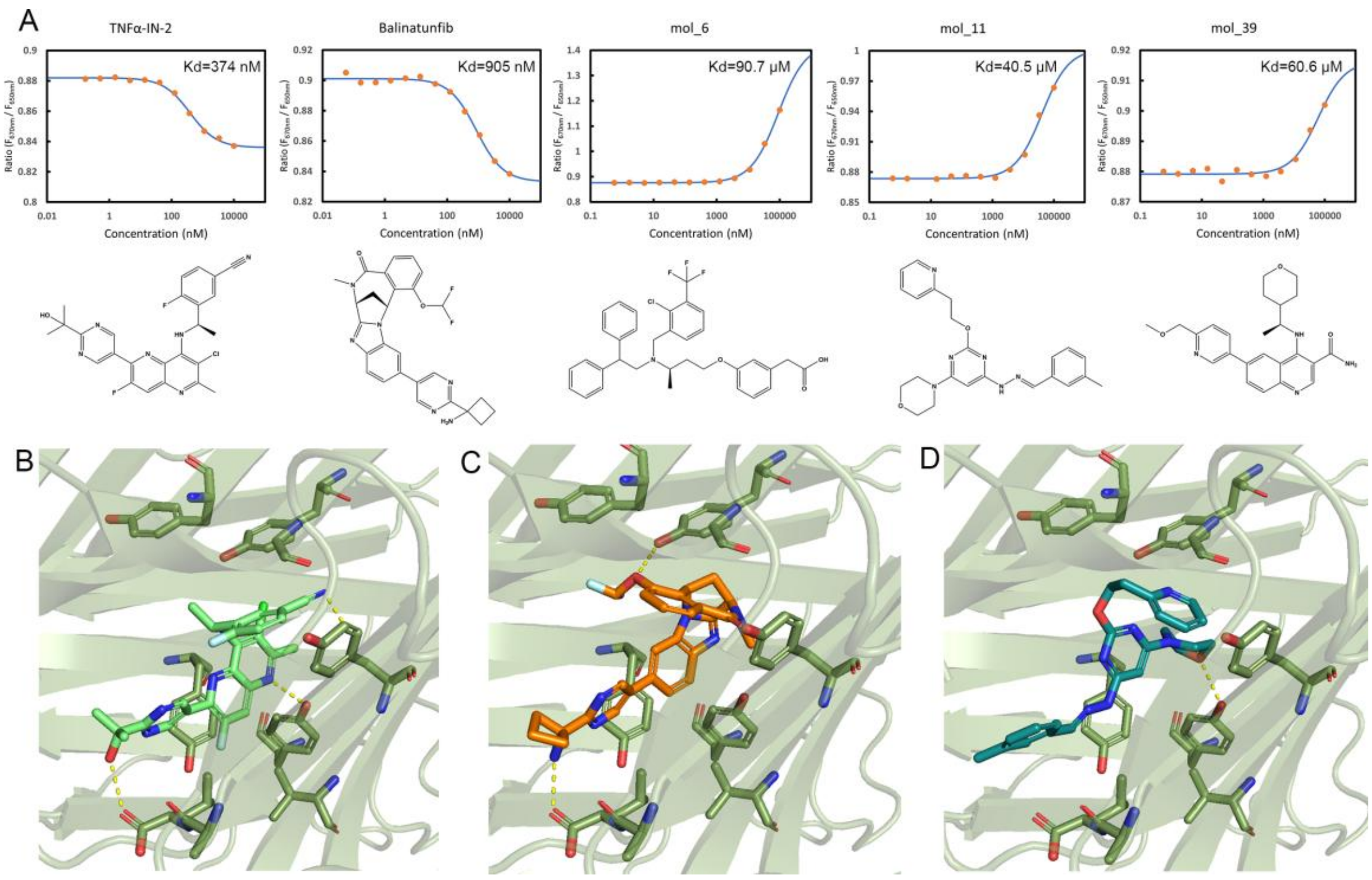

**Fig. 5:** (A) Binding curves of TNF-α-IN-2, Balinatunfib and compounds mol_6, mol_11 and mol_39. (B) Binding mode of TNF-α-IN-2 within TNFα generated by *TRIDS*. (C) Binding mode of Balinatunfib within TNFα generated by *TRIDS*. (D) Binding mode compound mol_11 within TNFα generated by *TRIDS*.

## 4. Discussion

The key to improving the accuracy of molecular docking lies in an accurate SF. Although recent ML-based methods have significantly improved the accuracy of SFs (e.g. RTMScore[27], GNINA[26]), most of them are subject to the docking-then-rescoring workflow, rendering the overall docking process cumbersome. It is desirable for the optimal conformation to be obtained through the global maximization of the SF, unifying the process of sampling and scoring. In the previous work of *DSDP*[16], we have demonstrated that powerful docking performance can be achieved when explicit gradient of analytic SF is used. We found in this study that certain properties are required to make the ML SFs suitable for guiding the conformation sampling. Firstly, the SF should be differentiable with respect to the coordinates of all the atoms of input ligands. Secondly, the SF should be as smooth and concise as possible for efficient calculation of gradients. This requirement excludes the classical ML methods (*e.g.* RF-Score[25]) or descriptor-based deep learning methods from being possible candidates.

Currently, there are two different strategies for constructing differentiable ML SFs: 1) In unsupervised learning methods (*e.g.* RTMScore[27]), maximum entropy is exploited as the loss function to train neural networks and to receive features of both receptors and ligands. They output the critical coefficients to construct mixed Gaussian models, which were then used to calculate the score based on the distance matrix between the atoms of ligands and residues of receptors. This distance-dependent strategy was introduced by DeepDock[46], without using the derivative of SF. 2) In supervised learning methods (*e.g.* GNINA score[26]), cross entropy or minimum square error was taken as the loss function to train neural networks, which takes coordinates and features of ligands and receptors as the input and the score value as the output. Comprehensive comparison showed that unsupervised learning suits better for conformation sampling. Co-crystal structures in public database such as PDBbind is only viewed as positive samples. The negative samples (decoy) are normally constructed manually based on self-defined criteria. These decoy data are essential for the training of supervised learning methods, but not being required for unsupervised learning methods. After obtaining the profiles predicted by the neural networks based on given ligands and receptors, not the entire neural network but only the weighted Gaussian terms are needed for differentiation. In supervised learning methods, the total neural network should be included in gradient computation. Furthermore, the gradient calculated on the limited Gaussian kernels is smoother than that calculated on the entire neural networks. Therefore, distance-dependent MDN method was chosen here as SF as well as for conformation sampling.

From physical point of view, MDN method could be regarded as inverse Boltzmann conversion, *i.e.*, the specific pairwise distance distribution could be inferred from systematic energy distribution. For the continuous system, the probability of states of the ligand-receptor complex could be expressed based on Boltzmann distributions as

follows:

$$p(x) = \frac{exp(-E(x)/kT)}{Z}$$

$$Z = \int_{-\infty}^{+\infty} exp(-E(x)/kT)dx$$

where x denotes these states, commonly described as the distance matrix derived from the atom coordinates. Z is partition function ensuring the normalization:

$$\int_{-\infty}^{+\infty} p(x)\, dx = 1$$

In the maximum likelihood estimation, it is a common hypothesis that the state from experimental co-crystal structure has the lowest energy and the largest probability to be observed. Therefore, the objective function could be defined as follows:

$$\nabla_x \mathcal{L}_{MLE} = \nabla_x \mathbb{E}_{p(x)} \left[ \log \frac{exp(-E(x)/kT)}{Z} \right]$$

$$= -\nabla_x \mathbb{E}_{p(x)}[E(x)] - \nabla_x \log \int_{-\infty}^{+\infty} exp(-E(x))dx$$

$$\approx -\nabla_x \mathbb{E}_{p(x)}[E(x)]$$

In the above equation the first term could be estimated from 3D structure of complexes, while the second term is intractable, but regarded as the constant. If the energy function is displaced with MDN model, it learns Boltzmann distribution with respect to the distance between pairwise atoms. As a proof, the distribution of the hydrogen bond, salt bridge and hydrophobic interaction obtained from the predicted ligand conformation have almost the same peak values of pairwise distances as in the co-crystal structures for both Time-split (Fig. 6A-C) and PoseBusters (Fig. 6D-F).

However, not all important physical knowledge can be distilled from static 3D structures. Especially, unwanted atomic clashes frequently occur between the ligand and receptor, possibly due to the scarcity in negative samples to provide enough clash information in the training set. In addition, long-term interactions are also difficult to be captured by MDN model alone. When critical atom pairs are separated by more than 5Å in the transient conformation, MDN model fail to provide effective gradient for optimization. Furthermore, since the existing MDN is a coarse-grained model, it cannot accurately recognize the formation of hydrogen bond during the sampling process.

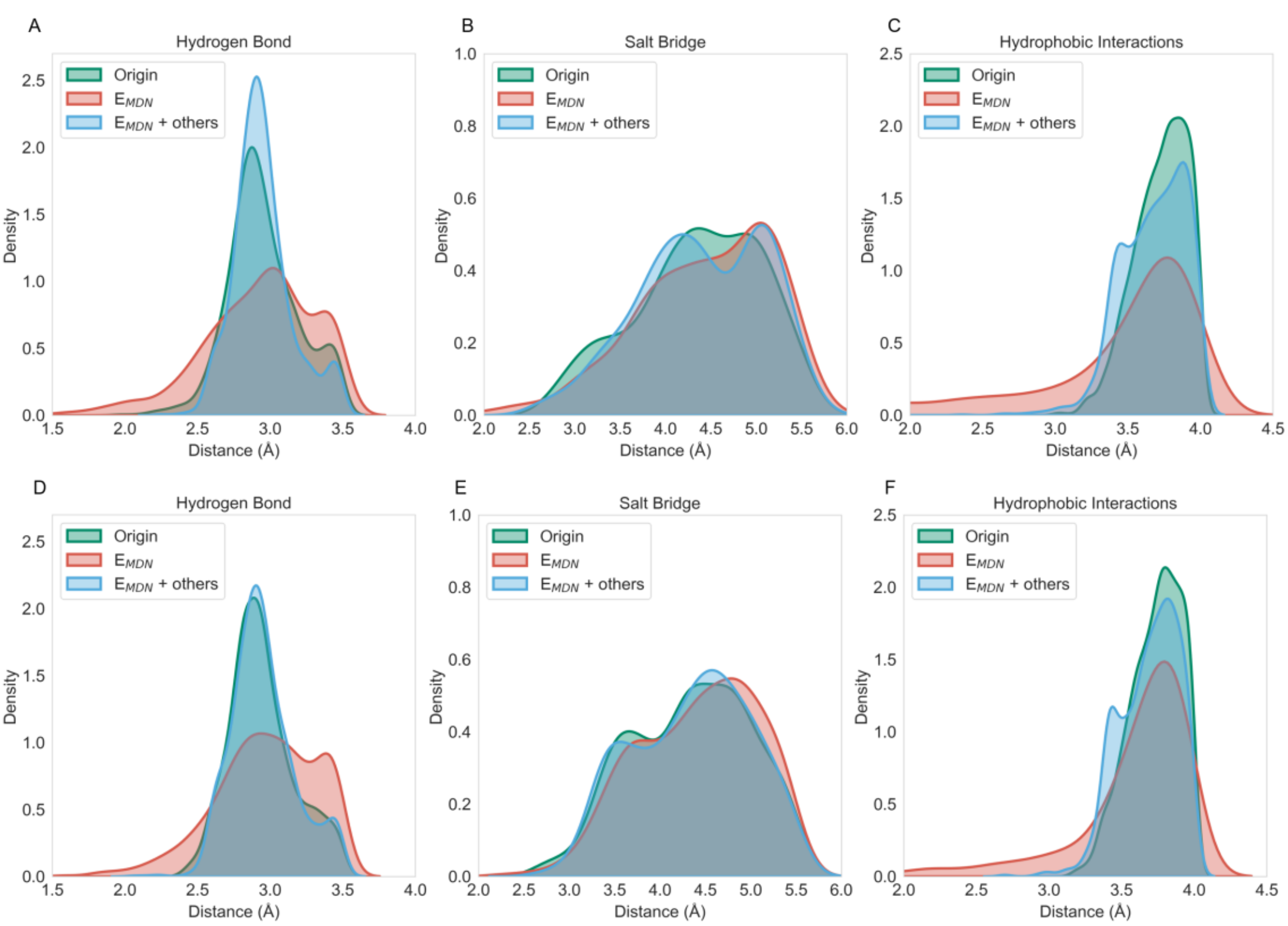

**Fig. 6:** The distance distributions of pairwsie atoms for three types of physics-informed SF terms (hydrogen bonds, salt bridges, and hydrophobic interactions) of the binding poses in crystallized structures as reference (Origin: green) and predicted by MDN-bsed *TRIDS* with ($E_{MDN}$ + *others*: blue) and without ($E_{MDN}$: red) revisions of physics-informed SF terms in Time-split (A-C) and PoseBusters (D-F).

To overcome these defects, we revised the model by adding other physics-informed terms representing long-term interaction, clash and hydrogen bond. MDN model can easily incorporate these terms through linear combination of the distance-based formula. This revision increased the docking accuracy by about 4%. In addition, the PB-valid rate was improved significantly and the distribution of generated critical atomic interactions was much closer to the original distribution for both Time-split (Fig. 6A-C) and PoseBusters (Fig. 6D-F). It is worth noting that the docking performance could be improved significantly with the inclusion of water forming hydrogen bonds (from 77.34% to 84.50% in PoseBusters), although the screening power slightly decays (Table S8 and S9).

Considering the integration of deep learning model and automatic differentiation in conformation sampling, PyTorch C++ is adopted to implement the method on both GPU and CPU. During conformation refinement, *TRIDS* runs a large number of copies to make full use of the parallel computational cores on GPUs in the sampling process, which allows it to handle relatively challenging docking problems, e.g., large ligands. Moreover, *TRIDS* is further optimized in the following aspects: 1) **Multiple stream concurrency**: because of asynchronized calculation between CPU and GPU, CUDA streams play a similar role as multi-thread in CPU, which allows multiple conformations to be optimized simultaneously with shared memory. 2) **Operator fusion in CUDA graph**: in order to save the communication time between GPU and CPU, the repetitive operators in the process of conformation sampling are compiled as a CUDA graph and saved on GPU memory in advance. 3) **Manual differentiation**: although PyTorch provides a comprehensive automatic differentiation for most commonly used operators, it sacrifices efficiency to achieve the best robustness. For computational bottlenecks such as Gaussian-based SFs and coordinate update based on degrees of freedom, we manually defined the operators with analytical gradient formula. 4) **Self-defined kernel function**: to avoid calculation for atom pairs of large distances, we directly defined the CUDA kernel function for the calculation of SF to only calculate atom-residue pairs within a given distance. As a result of these optimizations, *TRIDS* is able to deal with tens of thousands of binding poses in parallel with significantly reduced consumption of computational resources. Although its performance stemmed from coarse-grained MDN is not comparable with other AI-based docking methods for smaller molecules (<400 Da), its high efficiency of conformation sampling leads to outstanding performance for large compounds (Fig. 3C and D).

## 5. Conclusion

AI-powered methods have shown its potential to enhance the accuracy of molecular docking. However, existing methods often overlook the physical fitness of the docked complexes. In this work, we introduce *TRIDS*, an AI-native docking framework that unifies conformation sampling and dokcing processes with DL-based model to improve accuracy of docking and screening. Furthermore, it is tailored for seamless cooperation of AI and physics to guarantee the physical validity of predicted binding poses. As a result, the majority of the sampled conformations are accurate and consistent with physical rules. Notably, *TRIDS* achieves high computational efficiency, delivering conformation prediction in a fraction of second while maintaining a highly lightweight GPU memory footprint of merely hundreds of megabytes. Its performance was further test by wet-lab experiment in which hit compounds with novel scaffolds were obtained for tumor necrosis factor-alpha (TNFα) through high-throughput VS.

## Supporting Information

The following files are available along with this paper. Input node and edge features for the ligand and protein graphs, rescoring performance of MDN-based SF, SRs of *TRIDS* with and without water passing different criteria in the PoseBusters benchmark, hyper-parameters optimization, table of performance of different optimizers in Time-split, performance of blind docking, effect of retained water types on docking and screening performance are shown in the Supporting Information. (PDF)

## Data and software availability

a. *TRIDS* is available at https://github.com/xuhanliu/trids/ or https://gitee.com/xuhanliu/trids/.
b. The PDBbind (v2020) and CASF-2016 benchmark are available at https://www.pdbbind-plus.org.cn/.

c. BioLip is available at https://zhanggroup.org/BioLiP/.

d. DEKOIS 2.0 is available at http://www.dekois.com.

e. PoseBusters is available at https://zenodo.org/records/8278563.

## Author contributions

X.L. and B.Z. contributed equally. Y.G., H.Z., X.L. and L.Y. conceived the research project. X.L. developed the algorithm and implemented the program. B.Z., H.Z. and Y.H. estimated the program and undertook the benchmark. Y.X., T.B., L.C. and T.Z. contributed to the biological experiments. X.L., B.Z. H.Z. and Y.G. wrote the paper. All authors read and approved the manuscript.

## Declarations

The authors declare no competing interests.

## Acknowledgement

This work was supported by the National Natural Science Foundation of China T2495221, and the New Cornerstone Science Foundation NCI202305.

# Supporting information

## S1 Input features of node and edge for the ligand and protein target graphs

**Table S1:** Node and edge features employed for ligand graph construction

| Features | Size | Description |
|---|---|---|
| | | **Nodes (Atom)** |
| **Atom type** | 17 | one hot encoding for atom type ("C", "N", "O", "S", "F", "P", "CI", "Br", "T", "B", "Si", "Fe", "Zn", "Cu", "Mn", "Mo", "other") |
| **Explicit degree** | 7 | one hot encoding for atom degree (0, 1, 2, 3, 4, 5, 6) |
| **Formal charge** | 1 | formal charge |
| **Radical electrons** | 1 | number of radical electrons |
| **Hybridization** | 6 | one hot encoding for atom hybridization ("sp", "sp2", "sp3", "sp3d", "sp3d2", "other") |
| **Is aromatic** | 1 | whether the atom is aromatic |
| **Implicit Hydrogen** | 5 | one hot encoding for total number of Hs on the atom (0, 1, 2, 3, 4) |
| **Chirality** | 3 | one hot encoding for chirality of an atom ("R", "S", "other") |
| | | **Edges (Bond)** |
| **Bond type** | 4 | one hot encoding for bond type ("SINGLE", "DOUBLE", "TRIPLE", "AROMATIC") |
| **Is in ring** | 1 | whether the bond is in a ring |
| **Stereoisomer** | 4 | one hot encoding for the stereo configuration of a bond ("Shape_U", "Shape_4", "Shape_Z", "None") |

**Table S2:** Node and edge features employed for protein target graph construction

| Features | Size | Description |
|---|---|---|
| | | **Nodes (Residue)** |
| **Residue type** | 32 | one hot encoding for residue type ("GLY","ALA","VAL","LEU","LE","PRO","PHE","TYR","TRP","SER", "THR","CYS","MET","ASN","GLN","ASP","GLU","LYS","ARG","HIS" ,"MSE","CSO","PTR","TPO","KCX","CSD","SEP","MLY", "PCA","LLP","metal","other") |
| **Atomic self-distance** | 5 | maximum and minimum of the scaled distance (multiplied by 0.1) within any atom in a residue,<br>the scaled distance (multiplied by 0.1) between the atoms of CA and O<br>the scaled distance (multiplied by 0.1) between the atoms of O and N<br>the scaled distance (multiplied by 0.1) between the atoms of C and N |
| **Dihedral angle** | 4 | scaled angles (multiplied by 0.01), including phi, psi, omega and chil |
| | | **Edges (Residue-Residue pair)** |
| **Is connected** | 1 | whether two residues are connected by covalent bond |
| **CA distance** | 1 | scaled distance (multiplied by 0.1) between the CA atoms of two residues |
| **Center distance** | 1 | scaled distance (multiplied by 0.1) between the center of two residues |
| **Maximum distance** | 2 | maximum and minimum of the scaled distance (multiplied by 0.1) between two residues |

## S2 Docking and screening performance of MDN-based SF

The first step in the pipeline of *MDN* was to train an SF that achieved high performance of docking and screening. On the basis of RTMScore[1], we replaced RDkit with OpenBabel since the latter is much easier to be integrated into C++ programming. A MDN-based SF was then trained for molecular docking and VS tasks. Table S3 summarizes the performance of pretraining model, redocking model and screening model on CSAF-2016 benchmark. The docking SR (Fig. S1A), forward screening SR (Fig. S1B) and EF (Fig. S1C), and reverse screening SR of *TRIDS* was comparable to the performance of RTMScore.

**Table S3:** Docking and screening performance of three models on CASF-2016

| Model | Docking | Screening | | |
|---|---|---|---|---|
| | SR | Forward SR | Forward EF | Reverse SR |
| Model 1 | 94.40% | 70.20% | **28.33** | **38.20%** |
| Model 2 | **97.10%** | 56.10% | 26.08 | 36.80% |
| Model 3 | 95.42% | **64.90%** | 25.66 | 37.20% |

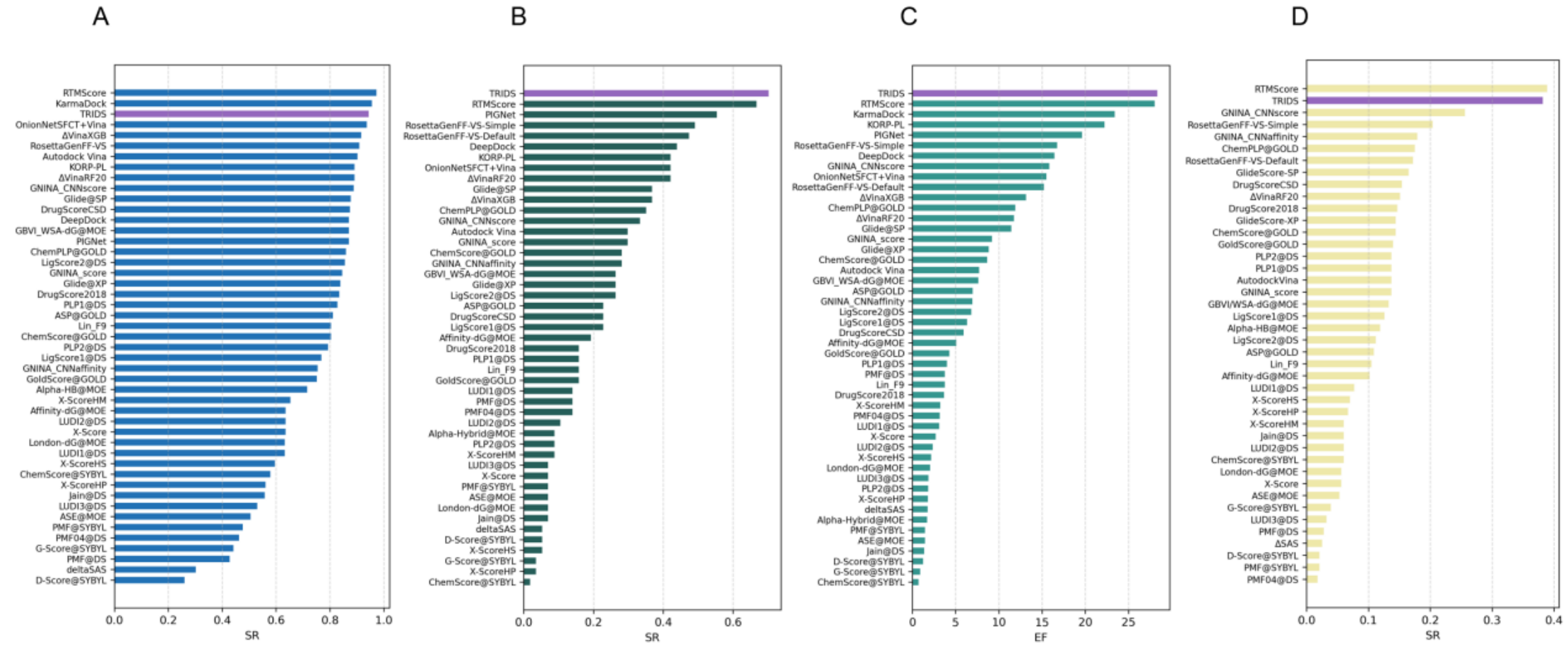

**Fig. S1:** Performances of *TRIDS* and other SFs evaluated on CASF-2016 (purple). The results of other methods including the SR of docking powers (A, blue) , forward screening power in terms of SR (B, green) and EF (C, cyan), and SR of the reverse screening (D, yellow) are from the work of RosettaGenFF[2,3].

## S3 SRs of *TRIDS* with and without water passing different PB-valid criteria

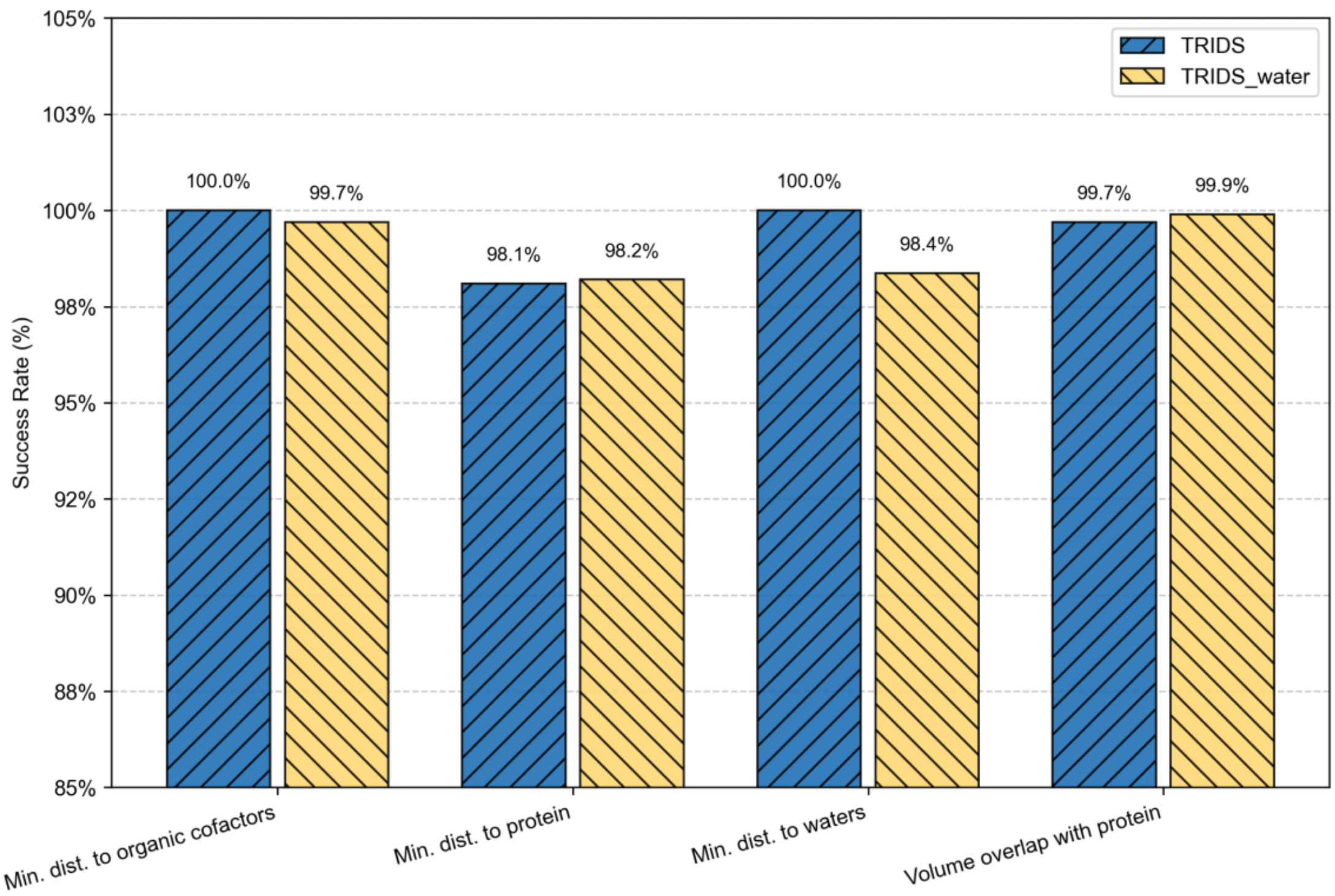

**Fig. S2**: SRs of *TRIDS* passing the different critera for receptors with (blue) and without (yellow) water on PoseBusters. The indicators that passed the two tests are not displayed here.

## S4 Hyper-parameters optimization for different SF terms in *TRIDS*

The weights of the physics-informed SF terms, including clash ($w_{clash}$), long-range interactions ($w_{long}$), hydrogen bonds ($w_{Hbond}$) and number of rotatable bonds ($w_{tor}$), and the type of optimizer for gradient descent in *TRIDS* were optimized stepwise as hyper-parameters on Time-split for docking tasks. The test results are shown in Table S4-6. The optimized weights were: $w_{clash}$ = 10, $w_{long}$ = 0.2, and $w_{Hbond}$ = 0.6. And Adam obtained the best performance among these optimizers.

**Table S4:** Optimizaton of weights of different physics-informed SF term combined with MDN-based SF for docking with *TRIDS*

| Weight types | Weight values | 1 | | | 2 | | | 3 | | | Average | | |
|---|---|---|---|---|---|---|---|---|---|---|---|---|---|
| | | Top-1 RMSD | | | Top-1 RMSD | | | Top-1 RMSD | | | Top-1 RMSD | | |
| | | <2Å (%)↑ | Med. (Å)↓ | <2Å & PB_valid | <2Å (%)↑ | Med. (Å)↓ | <2Å & PB_valid | <2Å (%)↑ | Med. (Å)↓ | <2Å & PB_valid | <2Å (%)↑ | Med. (Å)↓ | <2Å & PB_valid |
| $w_{clash}$ in $E_{clash}$ | 0 | 40.33 | 2.99 | 18.78 | 38.12 | 2.97 | 20.44 | 40.61 | 2.89 | 22.40 | 39.69 | 2.95 | 20.54 |
| | 1 | 56.91 | 1.57 | 40.06 | 57.46 | 1.54 | 39.5 | 57.46 | 1.58 | 40.88 | 57.28 | 1.56 | 40.15 |
| | 5 | 58.29 | 1.44 | 48.34 | 59.12 | 1.49 | 50.00 | 59.67 | 1.46 | 49.72 | 59.03 | 1.46 | 49.35 |
| | 10 | 58.84 | 1.51 | 51.38 | 59.67 | 1.34 | 51.93 | 59.94 | 1.44 | 51.93 | 59.48 | 1.43 | 51.75 |
| | 15 | 59.12 | 1.51 | 52.21 | 60.22 | 1.45 | 52.49 | 59.12 | 1.47 | 50.83 | 59.49 | 1.48 | 51.84 |
| | 20 | 55.80 | 1.69 | 49.72 | 56.35 | 1.66 | 50.28 | 58.66 | 1.55 | 51.66 | 56.94 | 1.63 | 50.55 |
| $w_{long}$ in $E_{long}$ | 0.0 | 58.84 | 1.51 | 51.38 | 59.67 | 1.34 | 51.93 | 59.94 | 1.44 | 51.93 | 59.48 | 1.43 | 51.75 |
| | 0.1 | 64.92 | 1.28 | 56.63 | 64.92 | 1.28 | 56.35 | 64.92 | 1.31 | 56.35 | 64.92 | 1.29 | 56.44 |
| | 0.2 | 64.64 | 1.31 | 56.08 | 65.75 | 1.24 | 55.80 | 65.75 | 1.31 | 56.63 | 65.38 | 1.29 | 56.17 |
| | 0.4 | 64.36 | 1.40 | 54.70 | 64.36 | 1.44 | 55.80 | 63.54 | 1.46 | 54.42 | 64.09 | 1.43 | 54.97 |
| | 0.6 | 60.22 | 1.59 | 51.38 | 59.67 | 1.60 | 51.38 | 59.94 | 1.54 | 50.83 | 59.94 | 1.58 | 51.20 |
| | 0.8 | 55.25 | 1.78 | 45.30 | 54.14 | 1.78 | 45.03 | 55.25 | 1.76 | 45.30 | 54.88 | 1.77 | 45.21 |
| $w_{Hbond}$ in $E_{Hbond}$ | 0.0 | 64.64 | 1.31 | 56.08 | 65.75 | 1.24 | 55.80 | 65.75 | 1.31 | 56.63 | 65.38 | 1.29 | 56.17 |
| | 0.2 | 64.64 | 1.24 | 56.35 | 67.68 | 1.23 | 58.29 | 65.19 | 1.27 | 55.25 | 65.84 | 1.25 | 56.63 |
| | 0.4 | 64.09 | 1.27 | 55.80 | 65.75 | 1.24 | 56.63 | 65.74 | 1.23 | 56.35 | 65.19 | 1.25 | 56.26 |
| | 0.6 | 66.02 | 1.21 | 56.91 | 67.4 | 1.24 | 58.84 | 67.4 | 1.25 | 57.73 | 66.94 | 1.23 | 57.83 |
| | 0.8 | 66.02 | 1.23 | 57.18 | 65.47 | 1.27 | 56.63 | 66.02 | 1.23 | 57.18 | 65.84 | 1.24 | 57.00 |
| | 1.0 | 66.85 | 1.23 | 57.73 | 64.09 | 1.23 | 55.52 | 65.47 | 1.25 | 55.8 | 65.47 | 1.24 | 56.35 |

**Table S5:** Docking performance of *TRIDS* with different optimizers for gradient descent on Time-split

| Optimizers | Adam | AdamW | RMSprop | AdaGrad | SGD |
|---|---|---|---|---|---|
| **SR (%)** | **66.94** | 58.4 | 63.36 | 60.33 | 2.2 |

The weight of the revision term with respect of rotatable bond numbers in *TRIDS* were optimized as hyperparameters on DEKOIS 2.0 for VS tasks (Table S5). Finally, the optimized weight is 0.01.

**Table S6:** Performance of *TRIDS* with different weights of revision term based on rotatable bond numbers in *TRIDS* for VS tasks on DEKOIS 2.0

| $w_{tor}$ | ROC-AUC | EF0.5% | EF1% | QED0.5% | QED1% |
|---|---|---|---|---|---|
| 0 | 0.814 | 21.493 | 19.388 | 45.10 | 46.40 |
| 0.005 | 0.813 | 21.429 | 19.420 | 46.10 | 46.80 |
| 0.01 | 0.813 | 21.365 | 19.484 | 47.00 | 47.50 |
| 0.02 | 0.811 | 21.235 | 19.292 | 48.00 | 48.60 |
| 0.03 | 0.808 | 21.172 | 19.069 | 48.80 | 49.60 |
| 0.04 | 0.806 | 20.917 | 18.845 | 49.40 | 50.50 |
| 0.05 | 0.804 | 20.279 | 18.685 | 50.30 | 51.10 |

## S5 Blind docking performance of *TRIDS*

**Table S7:** Performance of blind docking of *TRIDS* compared with other blind docking methods on PoseBusters

| Method | SR (RMSD < 2 Å) | PB_Valid (RMSD < 2 Å) |
|---|---|---|
| EquiBind | 2.6% | 0.7% |
| TankBind | 15.0% | 2.6% |
| DiffDock | 37.9% | 13.6% |
| TRIDS | **42.0%** | **41.1%** |

## S6 Effect of different water types for docking and screening performance

To systematically evaluate how different categories of water molecules affect the performance of *TRIDS*, we classified water molecules at the protein-ligand interface into three types based on their hydrogen-bonding patterns: ligand-bound water (forming hydrogen bonds with the ligand only), protein-bound water (forming hydrogen bonds with the protein only), and bridging water (forming hydrogen bonds with both the protein and the ligand). Using this classification, we designed six water-retention scenarios: 1) no water (removing all water molecules); 2) ligand-bound water only; 3) bridging water only; (iv) protein-bound water only; 4) all water except ligand-bound water (i.e., retaining protein-bound water and bridging water); and 5) all water (retaining all water molecules). Docking accuracy was assessed on PoseBusters, and screening performance was evaluated on DEKOIS 2.0. The results under these six scenarios are discussed in detail below. As shown in Tables S6 and S7, retaining water molecules that form hydrogen bonds with the ligand dramatically improves docking performance, although screening performance is slightly lower than that in the water-free scenario. The possible reason is that these water molecules provide more precise shape. However, for different ligands, water molecules could not be determined to be displaced or not. It is caused by the limitation of *TRIDS* that the water molecule could not be moved during the sampling process. For the next version of *TRIDS*, we will add the function of multiple component docking methods to determine which molecule could be displaced by the ligand.

**Table S8:** Docking performance of *TRIDS* under different water-retention scenarios on PoseBusters

| Scenario | | SR of top1 | SR top1 & PB_valid |
|---|---|---|---|
| No water | 1 | 77.10% | 75.70% |
| | 2 | 77.57% | 76.40% |
| | 3 | 77.34% | 75.47% |
| | **average** | **77.34%** | **75.86%** |
| Ligand-bound water only | 1 | 84.35% | 80.14% |
| | 2 | 84.58% | 80.84% |
| | 3 | 84.58% | 80.61% |
| | **average** | **84.50%** | **80.53%** |
| Bridging water only | 1 | 81.31% | 78.27% |
| | 2 | 82.00% | 79.21% |
| | 3 | 82.24% | 79.21% |
| | **average** | **81.85%** | **78.90%** |
| Protein-bound water only | 1 | 81.31% | 79.44% |
| | 2 | 81.31% | 78.27% |
| | 3 | 81.54% | 78.50% |
| | **average** | **81.39%** | **78.74%** |
| All water except ligand-bound water | 1 | 76.86% | 75.47% |
| | 2 | 78.04% | 75.93% |
| | 3 | 78.27% | 77.57% |
| | **average** | **77.72%** | **76.32%** |
| All water | 1 | 85.51% | 82.24% |
| | 2 | 86.21% | 83.41% |
| | 3 | 85.51% | 82.24% |
| | **average** | **85.74%** | **82.63%** |

**Table S9:** Screening performance of *TRIDS* under different water-retention scenarios on DEKOIS 2.0

| Scenario | | ROC-AUC | EF0.5% | EF1% |
|---|---|---|---|---|
| No water | 1 | 0.813 | 21.365 | 19.484 |
| | 2 | 0.816 | 21.365 | 19.356 |
| | 3 | 0.816 | 21.174 | 19.452 |
| | **average** | **0.815** | **21.301** | **19.431** |
| Ligand-bound water only | 1 | 0.790 | 18.559 | 16.963 |
| | 2 | 0.790 | 18.622 | 17.155 |
| | 3 | 0.791 | 18.366 | 16.677 |
| | **average** | **0.790** | **18.516** | **16.932** |
| Bridging water only | 1 | 0.794 | 18.431 | 17.123 |
| | 2 | 0.790 | 18.557 | 16.836 |
| | 3 | 0.793 | 18.750 | 16.995 |
| | **average** | **0.792** | **18.579** | **16.985** |
| Protein-bound water only | 1 | 0.685 | 11.029 | 9.692 |
| | 2 | 0.687 | 11.413 | 9.802 |
| | 3 | 0.686 | 11.541 | 9.852 |
| | **average** | **0.686** | **11.328** | **9.782** |
| All water except ligand-bound water | 1 | 0.686 | 11.668 | 10.106 |
| | 2 | 0.687 | 11.734 | 10.488 |
| | 3 | 0.685 | 11.541 | 10.425 |
| | **average** | **0.686** | **11.648** | **10.340** |
| All water | 1 | 0.668 | 9.308 | 8.193 |
| | 2 | 0.665 | 9.563 | 8.161 |
| | 3 | 0.664 | 9.181 | 7.715 |
| | **average** | **0.666** | **9.351** | **8.023** |